# Relationship among locally maximally entanglable states, W states and hypergraph states under local unitary transformations

Ri Qu, Yi-ping Ma, Bo Wang, and Yan-ru Bao

*School of Computer Science and Technology, Tianjin University, Tianjin, 300072, China and*

*Tianjin Key Laboratory of Cognitive Computing and Application, Tianjin, 300072, China*

Kruszynska and Kraus [Phys. Rev. A 79, 052304 (2009)] have recently introduced the so-called locally maximally entanglable (LME) states of $n$ qubits which can be maximally entangled to local auxiliary qubits using controlled operations. We characterize the local entanglability of hypergraph states and W states using an approach in [Phys. Rev. A 79, 052304 (2009)]. We show that (i) all hypergraph states are LME; (ii) hypergraph states and LME states are not equivalent under local unitaries; (iii) a W state of $n$ qubits is not LME; and (iv) no hypergraph state of $n$ qubits can be converted into to the W state under local unitary transformations. Moreover, we also present an approach for encoding weighted hypergraphs into LME states.

## I. INTRODUCTION

The understanding of the subtle properties of multipartite entangled states [1] is at the very heart of quantum information theory [2]. But the ultimate goal to cope with the properties of arbitrary multipartite states is far from being reached. Therefore, several special classes of entangled states have been introduced and identified to be useful for certain tasks. For instance, any *graph state* [3] can be constructed on the basis of a (simple and undirected) graph. *Cluster states* [4] are known to server as a universal resource for quantum computing in one-way quantum computer. *GHZ states* and W states [5] occur in quantum communication. *Stabilizer states* [6] can be employed for quantum error correction to protect quantum states against decoherence in quantum computation.

It is important to identify the relationship among deferent classes of entangled states. Graph states can describe a large family of entangled states including cluster states, GHZ states, and stabilizer states. But graph states cannot represent all entangled states (for instance, W states), which motivates us to introduce new classes of entangled states. To go beyond graph states and still keep the appealing connection to graphs, Ref. [7] introduces an axiomatic framework for mapping graphs to quantum states of a suitable physical system, and extends this framework to directed graphs and weighted graphs. Several classes of multipartite entangled states, such as *qudit graph states* [8], *Gaussian cluster states* [9], *projected entangled pair states* [10], and *quantum random networks* [11], emerge from the axiomatic framework. In [12], we generalize the above axiomatic framework to encoding hypergraphs into so-called quantum hypergraph states.

It is known that hypergraph states include graph states [12], and graph states cannot describe W states. Then one may ask whether there exists a hypergraph state of $n$ qubits such that it is equivalent to a W state of n qubits under local unitary transformations. Ref. [13] shows that no hypergraph state of three qubits can be converted into a W state of three qubits by local operations and classical communication (LOCC). The main aim of this work is to answer the above question for $n$-qubit hypergraph states $(n \geq 4)$. For this, we will address the issue of characterizing the local entanglability [14] of hypergraph states and W states by means of an approach introduced in [14]. We will show that (i) any hypergraph state is locally maximally entanglable (LME) [14]; (ii) hypergraph states and LME states are not equivalent under local unitaries; and (iii) all W states are not LME. Tow results (i) and (iii) will implies that our answer about the above question is "no". Moreover, we will indicate how to encode weighted hypergraphs into LME states.

This paper is organized as follows. In Sec. II, we recall notations of hypergraphs, hypergraph states, trace decompositions, LME states, etc. In Sec. III, we show the relationship among hypergraph states and LME states. We also indicate how to encode weighted hypergraphs into LME states. In Sec. IV, we prove that all W states are not LME. In Sec. V, we show that no hypergraph state can be converted into to a W state under local unitary transformations. Section VI contains our conclusions.

## II. PRELIMINARIES

Let a *W state* of $n$ qubits be $|W_n\rangle \equiv \frac{1}{\sqrt{n}}(|00...01\rangle + |00...10\rangle + ... + |10...00\rangle)$. Denote the $2\times 2$ identity matrix by $I$ and let

$$X \equiv \begin{bmatrix} 0 & 1 \\ 1 & 0 \end{bmatrix}, \ Y \equiv \begin{bmatrix} 0 & -i \\ i & 0 \end{bmatrix}, \text{ and } Z \equiv \begin{bmatrix} 1 & 0 \\ 0 & -1 \end{bmatrix}. \tag{1}$$

Denote an operator $V$ acting on qubit $l$ by $V_l$ while $V^k$ denotes the $k$th power of the operator $V$ with $V^0 \equiv I$ for any operator $V$. Let $|\phi\rangle$ and $|\varphi\rangle$ be two pure states of $n$ qubits. We say that they are *LU equivalent* if there exist local unitary operators $\{U_l\}_{l=1,2,...,n}$ such that

$$|\phi\rangle = U_1 \otimes U_2 \otimes ... \otimes U_n |\varphi\rangle, \tag{2}$$

i.e., $|\phi\rangle$ and $|\varphi\rangle$ are equivalent under local unitary transformations.

Let $|\phi\rangle$ be an $n$-qubit state with single qubit reduced states $\{\rho_l \equiv \text{Tr}_{\text{all but } l}(|\phi\rangle\langle\phi|)\}_{l=1,2,...,n}$. For any $l$, we can write the spectral decomposition of $\rho_l$, i.e.,

$$\rho_l = U_l^\dagger D_l U_l \tag{3}$$

where $D_l = diag(\lambda_1^{(l)}, \lambda_2^{(l)})$ and $\lambda_1^{(l)} \geq \lambda_2^{(l)} \geq 0$. We call $U_1 \otimes U_2 \otimes ... \otimes U_n |\phi\rangle$ a *trace decomposition* of $|\phi\rangle$ [14].

Formally, a *hypergraph* is a pair $(V, E)$, where $V$ is the set of *vertices*, $E \subseteq \wp(V)$ is the set of *hyperedges* and $\wp(V)$ denotes the power set of the set $V$. Let $Z_k$ be the $2^k \times 2^k$

diagonal matrix which satisfies

$$(Z_k)_{jj} = \begin{cases} -1 & j = 2^k \\ 1 & others \end{cases} \quad (4)$$

where $k$ is a nonnegative integer. Suppose that $V = \{1, 2, ..., n\}$ and $e \subseteq V$. Then the $n$-qubit *hyperedge gate* $Z_e$ is defined as $Z_{|e|} \otimes I^{\otimes n-|e|}$ which means that $Z_{|e|}$ acts on the qubits in $e$ while the identity $I$ acts on the rest. An $n$-qubit *hypergraph state* $|g\rangle$ can be constructed by $g = (V, E)$ as follows. Each vertex labels a qubit (associated with a Hilbert space $\mathbb{C}^2$) initialized in $|+\rangle \equiv \frac{1}{\sqrt{2}}(|0\rangle + |1\rangle)$. The state $|g\rangle$ is obtained from the initial state $|+\rangle^{\otimes n}$ by applying the hyperedge gate $Z_e$ for each hyperedge $e \in E$, that is,

$$|g\rangle = \prod_{e \in E} Z_e |+\rangle^{\otimes n}. \quad (5)$$

Thus hypergraph states of $n$ qubits are corresponding to $(\mathbb{C}^2, |+\rangle, \{Z_k | 0 \leq k \leq n\})$ by the axiomatic approach while graph states are related with $(\mathbb{C}^2, |+\rangle, Z_2)$ [7, 12].

Let $|\psi\rangle$ be a pure state of $n$ qubits. These qubits are called *system* ones. For each system qubit $l$ one can introduce a local *auxiliary* one $l_a$ with the initial state $|+\rangle \equiv \frac{1}{\sqrt{2}}(|0\rangle + |1\rangle)$. Let $C_l = \sum_{j=0}^{1} U_l^j \otimes |j\rangle_{l_a} \langle j|$ where $U_l$ is a unitary operator acting on system qubit $l$ and $|j\rangle_{l_a} \langle j|$ is the projector acting on the auxiliary qubit $l_a$ attached to $l$. If there exist local control gates $\{C_l\}_{l=1,2,...,n}$ such that the state $C_1 \otimes C_2 \otimes ... \otimes C_n |\psi\rangle |+\rangle^{\otimes n}$ is a maximally entangled state between the system and the auxiliary systems, then the state $|\psi\rangle$ is called *locally maximally entanglable* (LME) [14].

### III. RELATIONSHIP BETWEEN HYPERGRAPH STATES AND LME STATES

In this section we discuss the local entanglability of hypergraph states. We show that all hypergraph states of $n$ qubits are of LME states. But all LME states are not equivalent to hypergraph states under local unitaries, i.e., there exists a LME state such that it is not LU equivalent to any hypergraph state.

*Proposition 1*. Any hypergraph state is LME.

*Proof.* It is known that real equally weighted states [15] are equivalent to hypergraph states [12]. In fact, let $V = \{1, 2, ..., n\}$ and define a mapping $c$ on $\wp(V)$ as

$$\forall e \subseteq V, c(e) = \begin{cases} 1 & e = \Phi \\ \prod_{k \in e} x_k & e \neq \Phi \end{cases}. \tag{6}$$

Then we can construct a *1-1* mapping $u$ between hypergraphs and Boolean functions which satisfies $\forall g = (V, E)$,

$$u_g(x_1, x_2, ..., x_n) = \bigoplus_{e \in E} c(e). \tag{7}$$

where $\oplus$ denotes the addition operator over $\mathbb{Z}_2$. Thus we have

$$|g\rangle = \prod_{e \in E} Z_e |+\rangle^{\otimes n} = \frac{1}{\sqrt{2^n}} \sum_{x=0}^{2^n-1} (-1)^{\bigoplus_{e \in E} c(e)} |x\rangle = \frac{1}{\sqrt{2^n}} \sum_{x=0}^{2^n-1} (-1)^{u_g(x)} |x\rangle \equiv |\psi_{u_g}\rangle \tag{8}$$

where $|\psi_{u_g}\rangle$ is just the real equally weighted state associate with the Boolean function $u_g$. Then Eq. (8) can be rewritten into

$$|g\rangle = \frac{1}{\sqrt{2^n}} \sum_{x=0}^{2^n-1} e^{iu_g(x)\pi} |x\rangle. \tag{9}$$

According to Thereom 2 in [14], the state $|g\rangle$ is LME. ∎

Ref. [14] discusses some applications of LME states. Since all hypergraph states are LME, any hypergraph state can be used to encode classical information locally like LME states. It can also be used to implement certain non-local unitary operations. In the following we prove that LME states and hypergraph states are not LU equivalent.

*Proposition 2.* There exists a LME state such that it is not LU equivalent to any hypergraph state.

*Proof.* Let an *n*-qubit state $|\psi_f\rangle$ be

$$\frac{1}{\sqrt{2^n}} \sum_{x=0}^{2^n-1} e^{if(x)\pi} |x\rangle \tag{10}$$

where *f* is a function form the set $\{0, 1, ..., 2^n - 1\}$ to the real set $\mathbb{R}$. Clearly, the state $|\psi_f\rangle$ is LME by Thereom 2 in [14]. In particular, if *f* is a Boolean function (i.e., there is a hypergraph *g* such that $f = u_g$), then $|\psi_f\rangle$ is of hypergraph states by (9). The density operator of $|\psi_f\rangle$ can been written into

$$|\psi_f\rangle \langle \psi_f| = \frac{1}{2^n} \sum_{j,k=0}^{1} |j\rangle_1 \langle k| \otimes \sum_{x,y=0}^{2^{n-1}-1} e^{i[f(j,x)-f(k,y)]\pi} |x\rangle \langle y|. \tag{11}$$

Thus the single qubit reduced state of the first qubit can be obtained by

$$\rho_1^f = \mathrm{Tr}_{\text{all but 1}}\left(|\psi_f\rangle\langle\psi_f|\right) = \begin{bmatrix} \frac{1}{2} & \frac{1}{2^n}\chi_f \\ \frac{1}{2^n}\chi_f^* & \frac{1}{2} \end{bmatrix} \quad (12)$$

where $\chi_f = \sum_{x=0}^{2^{n-1}-1} e^{i[f(0,x)-f(1,x)]\pi}$. It is clear for any hypergraph state $|g\rangle$ that $\chi_{u_g}$ is an integer. Now we construct a special LME state $|\psi_f\rangle$ in (10) by defining the function $f$ as follows.

$$f(x) = \begin{cases} \alpha & x = 0 \\ 0 & x \in \{1, 2, \ldots, 2^{n-1}-1\} \\ 0 & x \in \{2^{n-1}, 2^{n-1}+1, \ldots, 2^n-1\} \end{cases}, \quad (13)$$

that is,

$$f(j,x) = \begin{cases} \alpha & j=0, x=0 \\ 0 & j=0, x \in \{1,2,\ldots,2^{n-1}-1\} \\ 0 & j=1, x \in \{0,1,\ldots,2^{n-1}-1\} \end{cases} \quad (13')$$

where $\cos(\alpha\pi) = \frac{1}{2^n}$. Then it is clear that $\chi_f = (2^{n-1}-1+e^{i\alpha\pi})$. Thus by (12) we obtain

$$\det[\rho_1^f] = \frac{1}{4} - \frac{1}{4^n}\chi_f\chi_f^* \quad (14)$$

Since

$$\chi_f\chi_f^* = (2^{n-1}-1)^2 + 2 - \frac{1}{2^{n-1}} \quad (15)$$

is not an integer, it is clear for any hypergraph state $|g\rangle$ that $\det(\rho_1^f) \neq \det(\rho_1^{u_g})$. It is known that the local *entropic measures* [16] are invariant under local unitary operations. Thus the state $|\psi_f\rangle$ is not LU equivalent to any hypergraph state. ∎

The above two propositions motivate us to generalize hypergraph states to introduce the definition of weighted hypergraph states which are constructed by weighted hypergraphs. We also show that weighted hypergraph states are equivalent to LME states under local unitaries, which implies weighted hypergraph states can describe more entangled states than hypergraph states.

At first, let us recall the definition of weighted hypergraphs. A *weighted hypergraph* is a pair $(V, \Gamma)$, where $V$ is the vertex set and $\Gamma: \wp(V) \to \mathbb{R}$ is the *weighted function*. A hypergraph $(V, E)$ defined in Sec. II can be regarded as the weighted hypergraph $(V, \Gamma)$ where the

weighted function $\Gamma$ satisfies

$$\Gamma(e) = \begin{cases} 1 & e \in E \\ 0 & e \notin E \end{cases} \quad (16)$$

Next we define weight hyperedge gates, which is similar for hyperedge gates defined in Sec. II. Let $Z_k(\alpha)$ be the $2^k \times 2^k$ diagonal matrix which satisfies

$$[Z_k(\alpha)]_{jj} = \begin{cases} e^{i\pi\alpha} & j = 2^k \\ 1 & others \end{cases} \quad (17)$$

where $k$ is a nonnegative integer and $\alpha \in \mathbb{R}$. Suppose that $V = \{1, 2, ..., n\}$ and $e \subseteq V$. Then the $n$-qubit *weighted hyperedge gate* $Z_e[\Gamma(e)]$ is defined as $Z_{|e|}[\Gamma(e)] \otimes I^{\otimes n-|e|}$ which means that $Z_{|e|}[\Gamma(e)]$ acts on the qubits in $e$ while the identity $I$ acts on the rest. This means that $Z_e[\Gamma(e)]$ can be regarded as a generalized ($|e|$-body) Ising interaction. Thus an $n$-qubit *weighted hypergraph state* $|G\rangle$ can be constructed by $G = (V, \Gamma)$ as follows. Each vertex labels a qubit initialized in $|+\rangle$. The state $|G\rangle$ is obtained from the initial state $|+\rangle^{\otimes n}$ by applying $Z_e[\Gamma(e)]$ for each hyperedge $e \subseteq V$, that is,

$$|G\rangle = \prod_{e \subseteq V} Z_e[\Gamma(e)] |+\rangle^{\otimes n}. \quad (18)$$

Note that there exists a *1-1* correspondence between hypergraphs with $n$ vertices and hypergraph states of $n$ qubits while a weight hypergraph state of $n$ qubits can be constructed by some different weighted hypergraphs with $n$ vertices. In fact, suppose that $G = (V, \Gamma)$ and $G' = (V, \Gamma')$ are two weighed hypergraphs. They satisfy that for each $e \subseteq V$,

$$\Gamma'(e) = \Gamma(e) + 2k \quad (19)$$

where $k$ is some nonzero integer. According to (17) and (18), it is evident that $|G\rangle = |G'\rangle$ and $G \neq G'$. It is clear that (18) is just the form of (2) in Ref. [14], that is, weighted hypergraph states and LME states are LU equivalent.

## IV. RELATIONSHIP BETWEEN W STATES AND LME STATES

Now let us discuss the relationship between W states and LME states. We show that the W state $|W_n\rangle$ is not of LME states as follows.

*Proposition 3.* The W state $|W_n\rangle$ is not LME.

*Proof.* Assume that $|W_n\rangle$ is LME. According to Lemma 1 in Ref. [14], there exists for each qubit $l$ a unitary operation $U_l$ such that the set $\{U_1^{l_1} \otimes U_2^{l_2} \otimes ... \otimes U_n^{l_n} |W_n\rangle\}_{l_1,l_2,...,l_n=0,1}$ forms a normal orthogonal basis. Let $\rho_l \equiv \text{Tr}_{\text{all but } l}(|W_n\rangle\langle W_n|)$. It is clear that $\rho_l = diag\left(\frac{n-1}{n}, \frac{1}{n}\right)$, which implies $|W_n\rangle$ is one trace decomposition. Since $\rho_l \not\propto I$, there is a real number $\alpha_l$ such that

$$U_l = R_{Z_l}(\alpha_l) X_l R_{Z_l}(-\alpha_l) = \begin{bmatrix} 0 & e^{i\alpha_l} \\ e^{-i\alpha_l} & 0 \end{bmatrix} \quad (20)$$

where $R_{Z_l}(\alpha_l) \equiv e^{i\alpha_l Z_l/2}$ [14]. For any two qubits $j$ and $k$, we can obtain

$$\langle W_n | U_j \otimes U_k | W_n \rangle = \frac{2}{n} \cos(\alpha_j - \alpha_k). \quad (21)$$

It is impossible that $\cos(\alpha_j - \alpha_k) = 0$ for any two $j$ and $k$. In fact, assume that $\cos(\alpha_j - \alpha_k) = 0$ and $\cos(\alpha_k - \alpha_l) = 0$. Then we would obtain $|\cos(\alpha_j - \alpha_l)| = 1$. ∎

## V. RELATIONSHIP BETWEEN HYPERGRAPH STATES AND W STATES

The W state $|W_n\rangle$ is one of famous $n$-partite (genuinely) entangled pure states of $n$ qubits. It has been applied for several quantum information processing tasks. Thus the preparation of the W state is very important. Clearly, for $n \geq 3$ no graph state of $n$ qubits is LU equivalent to the W state. In fact, it is known that the graph state constructed by a disconnected graph with $n$ vertices is not equivalent to the state $|W_n\rangle$ since it is not $n$-partite (genuinely) entangled. Let $g$ be a connected graph with $n$ vertices. It is known that all single qubit reduced density matrices $\rho_l$ of $|g\rangle$ satisfy $\rho_l \propto I$ [17]. Moreover, for the state $|W_n\rangle$ all single qubit reduced density matrices $\rho_l \not\propto I$, which is shown in the proof of the above proposition. Thus the state $|W_n\rangle$ cannot be prepared by using graph states under local unitaries according to the properties of entropic measure. Now we discuss the problem of the preparation of the W state by means of hypergraph states. The following proposition shows that no hypergraph state of $n$ qubits can be converted into the state $|W_n\rangle$ under local unitary transformations.

*Proposition 4.* No hypergraph state of $n$ qubits is LU equivalent to the W state $|W_n\rangle$.

*Proof.* According to Proposition 1, any hypergraph state is LME and can be written into the form shown in (10). Moreover, the state $|W_n\rangle$ is not LME by Proposition 3. Then it is not LU equivalent to any state in (10) according to Thereom 2 in [14]. Thus no hypergraph state of $n$ qubits is LU equivalent to $|W_n\rangle$. ∎

Clearly, the state $|W_n\rangle$ cannot be prepared by weighted hypergraph states according to Sec. III and the proposition 3. Note that the W state of three qubits cannot be prepared by using hypergraph states under SLOCC [13]. For $n \geq 4$ the problem whether the state $|W_n\rangle$ can be prepared by hypergraph states under SLOCC is still open.

## VI. CONCLUSIONS

We study the properties of the local entanglability of hypergraph states and W states by using an approach presented in [14]. As shown in Fig. 1, we describe the relationship among hypergraph states, W states and LME states under local unitaries. All hypergraph states are LME, that is, LME states include hypergraph states. This implies that hypergraph states may be use for the same quantum information processing tasks as LME states. For instance, they can be used to encode classical information locally, and to implement certain non-local unitary operators. But there is a LME state such that it is not LU equivalent to any hypergraph state, that is, LME states and hypergraph states are not equivalent under local unitaries. Furthermore, we generalize hypergraph states to introduce the so-called weighted hypergraph states which are just equivalent to LME states under local unitraries. In particular, it is interesting that the state $|W_n\rangle$ cannot be converted into any hypergraph state of $n$ qubits under local unitary transformations.

## ACKNOWLEDGMENTS

This work is supported by the Chinese National Program on Key Basic Research Project (973 Program, Grant No. 2013CB329304) and the Natural Science Foundation of China (Grant Nos. 61170178, 61105072 and 61272265). This work is completed during our academic visiting at Department of Computing, Open University, UK.

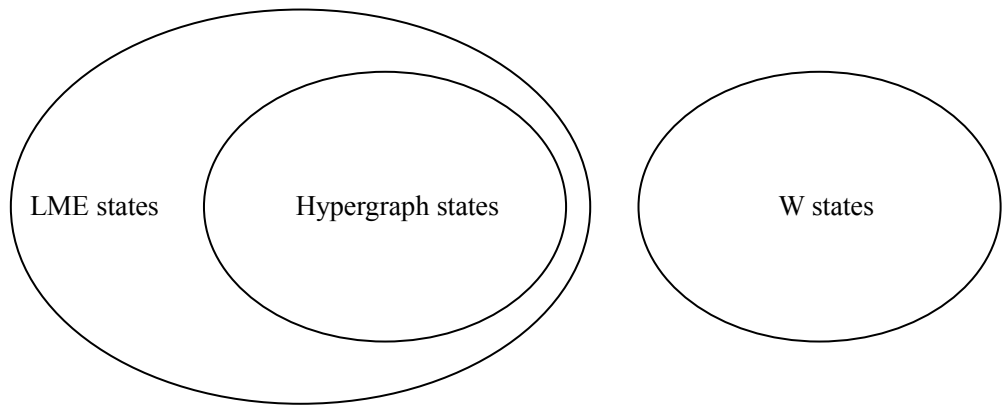

Figure 1. The relationship among LEM states, hypergraph states and W states under local unitary transformations.